\documentstyle[12pt,moriond,epsfig]{article}

\def\d0{D\O}
\def\D0{D\O}

\def\wg{$W\gamma$}
\def\ww{$WW$}

\def\wz{$WZ$}
\def\wwg{$WW\gamma$}
\def\wwv{$WWV$}
\def\wwz{$WWZ$}

\def\etmisv {\mbox{${\hbox{${\vec E}$\kern-0.6em\lower-.1ex\hbox{/}}}_T$}}
\def\etmis  {\mbox{${\hbox{$E$\kern-0.6em\lower-.1ex\hbox{/}}}_T$}}

\def\ifmath#1{\relax\ifmmode #1\else $#1$}%
\def\TeV{\ifmmode {\mathrm{ Te\kern -0.1em V}}\else
                   \textrm{Te\kern -0.1em V}\fi}%
\def\GeV{\ifmmode {\mathrm{ Ge\kern -0.1em V}}\else
                   \textrm{Ge\kern -0.1em V}\fi}%
\def\MeV{\ifmmode {\mathrm{ Me\kern -0.1em V}}\else
                   \textrm{Me\kern -0.1em V}\fi}%
\def\GeVcc{\ifmmode {\mathrm{ \GeV/c^2}}\else
                   \textrm{Ge\kern -0.1em V/c$^2$}\fi}%
\def\MeVcc{\ifmmode {\mathrm{ \MeV/c^2}}\else
                   \textrm{Me\kern -0.1em V/c$^2$}\fi}%

\def\geant{{\sc geant}}

\def\qwe{\alpha}

\begin{document}

\vskip -4cm 
\rightline{UCR/{D\O}/98-21}
\rightline{{D\O} Note 3450}
\rightline{FERMILAB-Conf-98/170-E}
\vskip 1cm

\vspace*{4cm}
\title{MEASUREMENTS OF THE W BOSON MASS AND TRILINEAR
GAUGE BOSON COUPLINGS AT THE TEVATRON}

\author{ JOHN ELLISON$^\dag$\footnotetext{$^\dag$For the CDF and \d0\ 
Collaborations}}

\address{Department of Physics, University of California, Riverside,\\
California 92521, USA}

\maketitle\abstracts{
We present measurements of the $W$ boson mass at the Tevatron based on $W
\to \mu \nu$ events collected by CDF and $W
\to e \nu$ events observed by \d0\ in Run Ib (1994--95). The $W$ boson mass 
measured in the preliminary CDF analysis is $80.43 \pm 0.10~{\mathrm
(stat)} \pm 0.12~{\mathrm (syst)}$~GeV$/c^2$. The \d0\ measured value
is $80.44 \pm 0.10~{\mathrm (stat)} \pm 0.07~{\mathrm
(syst)}$~GeV$/c^2$.  We also describe measurements of the trilinear
gauge boson couplings.  The limits obtained on the \wwg\ and \wwz\
anomalous couplings from a combined
\d0\ analysis using $W\gamma$, $WW \to \ell \nu
\ell^\prime \nu^\prime$, and $WW/WZ \to e \nu jj$ production are: 
$-0.30 < \Delta\kappa < 0.43$, $-0.20 < \lambda < 0.20$, and 
$-0.52 < \Delta g^Z_1 < 0.78$, for a dipole form factor scale of 2~TeV.
Improved limits have been obtained by combining these results with 
the limits derived from the LEP experiments.}

\begin{center}
{\it (Presented at the XXXIIId Rencontres de Moriond,
Electroweak Interactions and Unified Theories, Les Arcs, Savoie, 
France, March 14--21 1998.)}
\end{center}

\newpage

\section{Measurement of the W Boson Mass}

The mass of the $W$ boson is a fundamental parameter of the standard
model and is related to the Fermi constant $G_F$, the electromagnetic
coupling constant $\alpha_{EM}$, the $Z$ boson mass $m_Z$, and $\Delta
r$, which represents the effects of radiative corrections. $G_F$,
$\alpha_{EM}$, and $m_Z$ are all
measured\hspace{4pt}\cite{pdg}
with high precision. In the
standard model $\Delta r$ depends on the top quark and Higgs boson
masses and in theories beyond the standard model it depends on the
particle spectrum of the new theory. Therefore, together with
a measurement of the top mass, a precise measurement of the $W$ boson
mass can be used to constrain the Higgs mass in the standard model and
to constrain theories beyond the standard model.

The recent measurement published by {\d0}\hspace{4pt}\cite{d0_wmass_1b}
and the preliminary
measurement from CDF\hspace{4pt}\cite{cdf_wmass_1b}, 
both from Run Ib data (1994--95), are briefly
described here. The measurements are made using a fit to the observed
transverse mass spectrum $m_T = \sqrt{2p_T^\ell \etmis (1 - {\mathrm
cos} \Delta \phi)}$ in $W \to \ell \nu$ events.
The transverse mass spectrum is modeled using a Monte Carlo event generator
which incorporates a $W$ boson production model and a detailed model of the detector
response, which is calibrated using collider data.

\begin{figure}[p]
    \epsfysize = 5cm
    \centerline{\epsffile{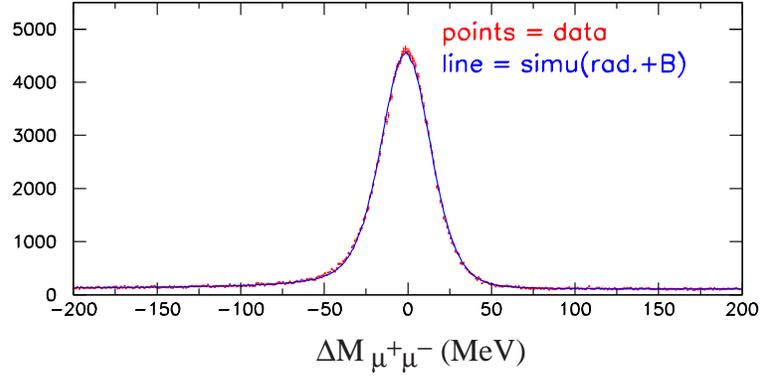}}
\caption{
Dimuom mass peak obtained from reconstructed $J/\psi \to \mu^+ \mu^-$
events in CDF. The points are the data and the line is the simulation,
which includes QED corrections and effects of
$B$-decays on the beam-constrained momentum measurement.}
\label{fig:cdf_jpsi}
\end{figure}

Calibration of the muon momentum scale is achieved in CDF by comparing
the reconstructed $J/\psi \to \mu^+ \mu^-$ mass
(Fig.~\ref{fig:cdf_jpsi}) to the world average.  The error in the $W$
boson mass due to the momentum scale uncertainty is $\delta m_W =
40$~MeV$/c^2$, while the momentum resolution contributes $\delta m_W =
25$~MeV$/c^2$.

\begin{figure}[p]
    \epsfysize = 6cm
    \centerline{\epsffile{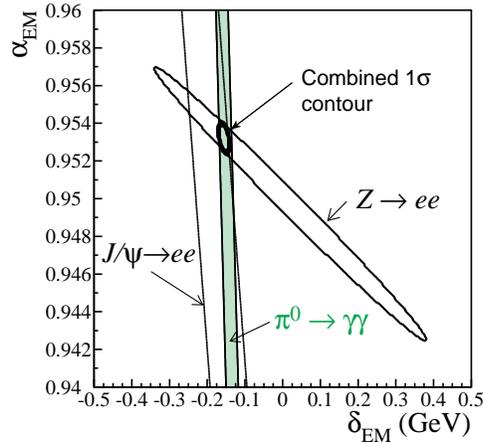}}
\caption{
Constraints on the parameters $\alpha$ and $\delta$ obtained in the \d0\
analysis using $\pi^0 \to \gamma \gamma$, $J/\psi \to ee$, and $Z \to ee$ events.}
\label{fig:d0_emscale}
\end{figure}

In \d0\ the electromagnetic calorimeter energy scale is determined
from test beam measurements and collider data. The observed energy
$E_{\mathrm obs}$ is parametrized as $E_{\mathrm obs} = \delta +
\alpha E_{\mathrm true}$, and the constants $\delta$ and $\alpha$ are
determined from $\pi^0 \to \gamma \gamma$, $J/\psi \to ee$, and $Z \to
ee$ events as shown in Fig.~\ref{fig:d0_emscale}. The resulting values
are $\alpha = 0.9533 \pm 0.0008$, and $\delta = (0.16 ^{+0.03}
_{-0.21})$~GeV, where the errors include the systematic uncertainty
due to underlying event corrections and non-linearity of the response
at low $E_T$. The contribution of the energy scale uncertainty to the
$W$ boson mass error is $\delta m_W = 70$~MeV$/c^2$. The energy resolution
contributes $\delta m_W = 25$~MeV$/c^2$.

\begin{figure}[p]
    \epsfysize = 5cm
    \centerline{\epsffile{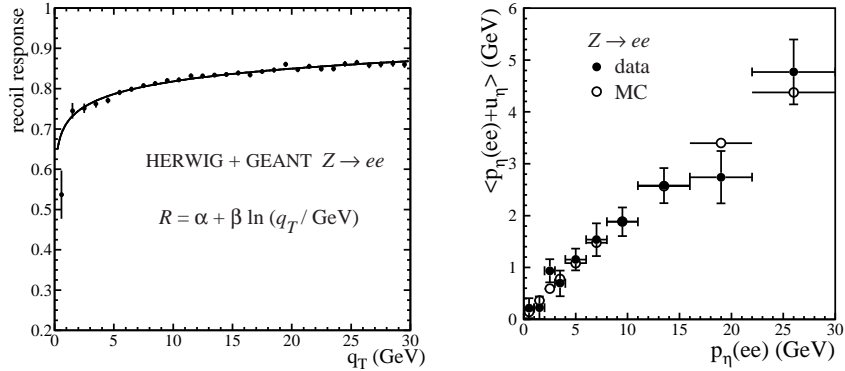}}
\caption{
Determination of the hadronic recoil response in \d0. 
(a) simulated recoil response $R$ versus $Z$ boson transverse 
momentum $q_T$. The points are from a detailed \geant\ simulation 
of the \d0\ detector and the line is the result of a fit using the function 
shown.
(b) $e^+e^-$-pair plus recoil system momentum 
$< p_\eta(ee) + u_\eta >$ versus the momentum of the $e^+e^-$-pair 
$p_\eta(ee)$. The quantities are projected along the $\eta$-axis, defined as 
the inner bisector of the $e^+$ and $e^-$ in the transverse plane. 
}
\label{fig:d0_recoil}
\end{figure}

In both CDF and \d0\ the response of the detector to the recoil
system, (hadrons recoiling against the $W$ boson, interactions of the proton
and antiproton spectator quarks, and energy from multiple
interactions), is calibrated using the transverse energy balance in $Z
\to ee$ decays.  The method employed by \d0\ is illustrated in
Fig.~\ref{fig:d0_recoil}.  The recoil response $R$ is defined by

$$|{\mathbf u}_T \cdot \hat {\mathbf q}_T| = R|{\mathbf q}_T|$$

\noindent
where ${\mathbf u}_T$ is the transverse momentum of the recoil system,
${\mathbf q}_T = q_T \hat {\mathbf q}_T$ is the transverse momentum of
the $Z$ boson. The LHS of this equation is the projection of the
recoil system transverse momentum along the $Z$ boson transverse
momentum vector, and for an ideal detector $R=1$.  A detailed {\geant}-based
Monte Carlo simulation shows that the response can be parametrized
using two constants $\alpha$ and $\beta$ 
(see Fig.~\ref{fig:d0_recoil}), which are determined using
$Z \to ee$ data, yielding $\alpha = 0.693 \pm 0.060$, and $\beta =
0.040 \pm 0.021$. The resulting contribution to the $W$ boson mass error is
$\delta m_W = 20$~MeV$/c^2$. The contribution from the recoil
resolution is $\delta m_W = 25$~MeV$/c^2$.

\begin{figure}[p]
    \epsfysize = 4.75cm
    \centerline{\epsffile{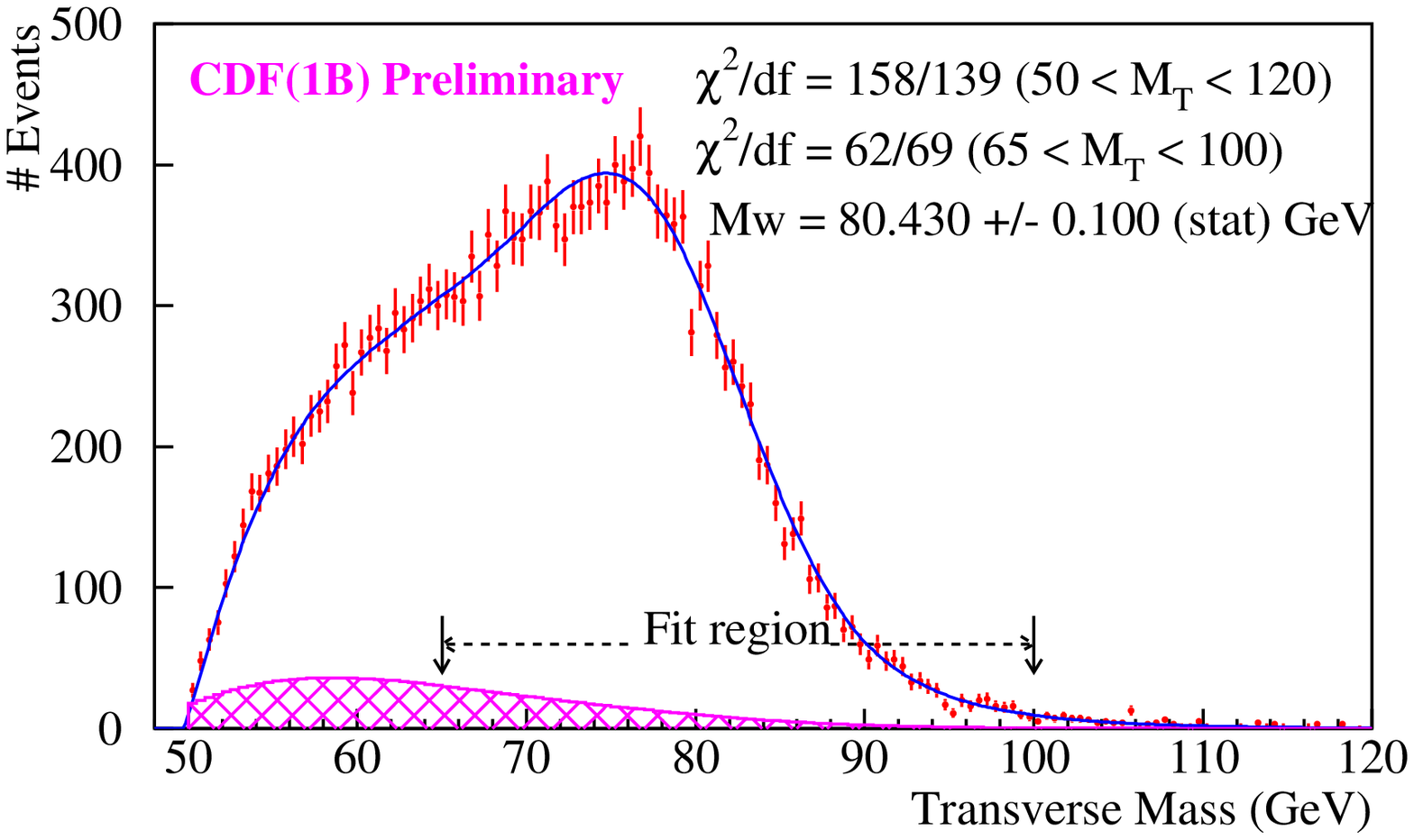}}
\caption{
Transverse mass distribution observed by CDF (points) and modeled 
by the Monte Carlo simulation for the best fit value of the $W$ boson mass (curve).
The contribution from the background is also shown (shaded distribution).
}
\label{fig:cdf_mtfit}
\end{figure}

\begin{figure}[p]
    \epsfysize = 5.5cm
    \centerline{\epsffile{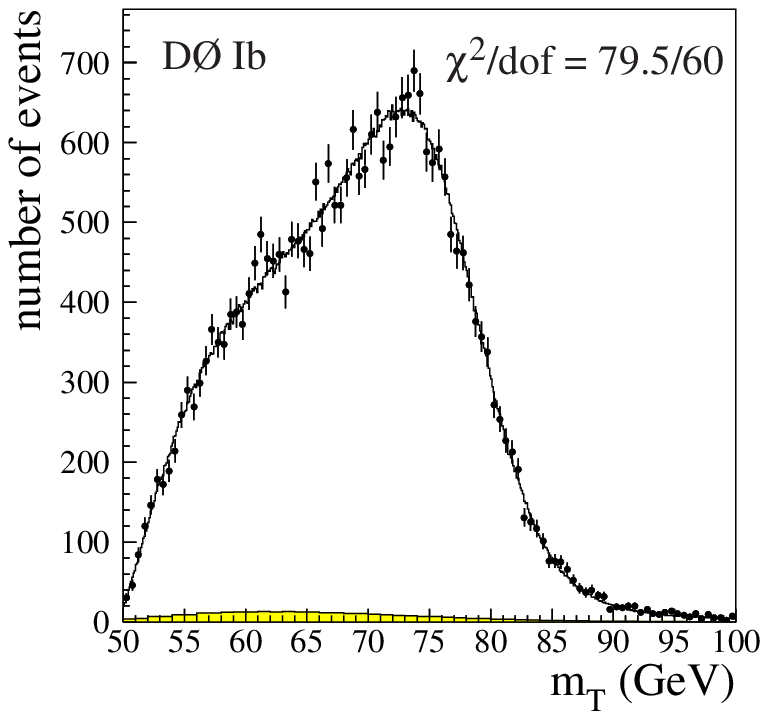}}
\caption{
Transverse mass distribution observed by \d0\ (points) and modeled 
by the Monte Carlo simulation for the best fit value of the $W$ boson mass (curve).
The contribution from the background is also shown (shaded distribution).
}
\label{fig:d0_mtfit}
\end{figure}

Fits to the transverse momentum distributions are shown in
Fig.~\ref{fig:cdf_mtfit} and Fig.~\ref{fig:d0_mtfit}. The CDF data
yield the result
$m_W = 80.43 \pm 0.10~{\mathrm (stat)} \pm 0.12~{\mathrm (syst)}$~GeV$/c^2$,
and the \d0\ result is 
$m_W = 80.44 \pm 0.10~{\mathrm (stat)} \pm 0.07~{\mathrm (syst)}$~GeV$/c^2$. 
Table~\ref{table:wmass} itemizes the sources of uncertainty in the
measurements.

\begin{figure}[p]
    \epsfysize = 6cm
    \epsfxsize = 8cm
    \centerline{\epsffile{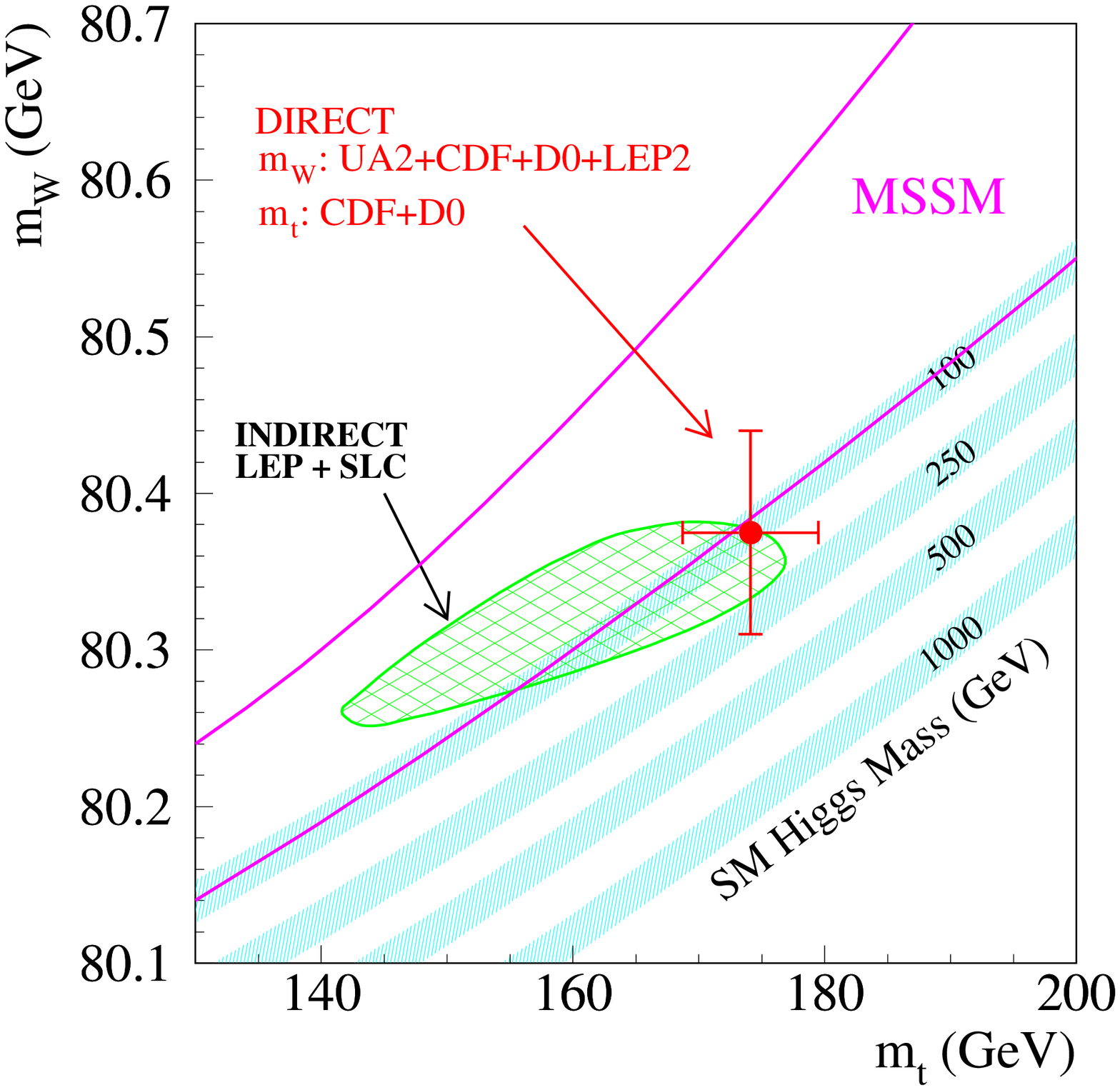}}
\caption{
$W$ boson mass $m_W$ plotted versus top mass $m_t$. The data point
represents the combined result from direct measurements.  The
shaded area is the allowed region from fits to the electroweak
prarameters.  Also shown are the standard model predictions for higgs
masses between 100$-$1000~GeV$/c^2$ and the prediction of the minimal
supersymmetric model (MSSM).}
\label{fig:mtmw}
\end{figure}

Combining with previous measurements\hspace{4pt}\cite{mw_previous} from UA2,
CDF and \d0\ yields a hadron
collider average of $m_W = 80.40 \pm 0.09$~GeV$/c^2$. The LEP
average $W$ boson mass reported at this conference\hspace{4pt}\cite{lep_wmass} is $m_W =
80.35 \pm 0.09$~GeV$/c^2$. Combining these results yields a new world
average of $m_W = 80.375 \pm 0.065$~GeV$/c^2$.
Combining this result with the Tevatron top mass 
measurement\hspace{4pt}\cite{top_mass}
($m_t = 174.1 \pm 5.4$~GeV$/c^2$) allows a comparison with the predictions of
the standard model\hspace{4pt}\cite{dr_sm} and the minimal supersymmetric 
model\hspace{4pt}\cite{dr_mssm},
as shown in Fig.~\ref{fig:mtmw}.

\begin{table*}[ht]
{\footnotesize
\begin{center}
\begin{tabular}{lcc}  \hline \hline
                   &\d0\ Run Ib   &CDF Run Ib prelim.\\
                   &[GeV$/c^2$]   &[GeV$/c^2$]\\ \hline

$W$ Statistics  & 70 & 100\\

& & \\

$E(e)$ or $p(\mu)$ scale  & 70 & 40\\

$e$ or $\mu$ resoultion	  & 40  & 25\\

Recoil modeling	  & 30  & 90\\

Selection bias    & $-$ & 20\\

\\

Backgrounds	  & 10 & 25\\

\\

$W$ width	  & 10 & 10\\

$W$ production (incl. pdf's) & 25 & 50\\

QCD / QED corrections & 15 & 30\\

\hline
\end{tabular}
\end{center}
}
\caption{
Contributions to the total $W$ boson mass uncertainty in the \d0\ and CDF 
Run Ib analyses.}
\label{table:wmass}
\end{table*}

\section{Trilinear Gauge Boson Couplings}

Gauge invariance under the group SU(2)~$\times$~U(1), an underlying
principle at the heart of the standard model, leads to the prediction
of gauge boson self-couplings (e.g. the \wwg\ and \wwz\ vertex
couplings). These couplings may be studied at the Tevatron through the
production of gauge boson pairs (\wg, \ww, \wz).  
Deviations from the
standard model would provide important information about the kind of
new physics beyond the standard model.

To test the agreement with the standard model and to set limits on
anomalous couplings the \wwv\ $(V=\gamma,Z)$ vertices are
parametrized using the effective Lagrangian of Ref.~9.
Assuming electromagnetic gauge
invariance, and invariance under Lorentz and CP transformations the
effective Lagrangian is reduced to a function of five dimensionless
coupling parameters $g_1^Z, \kappa_V$, and $\lambda_V$.  In the SM at
tree level $g_1^Z = 1, \Delta \kappa_V \equiv \kappa_V - 1 = 0$
and $\lambda_V = 0$.

%
%

The effective Lagrangian formalism is valid only at energies much
smaller than the scale of new physics. At very high energies the
formalism breaks down and the full particle spectrum of the new theory
must be included to ensure unitarization. In the hadron collider
experiments it is customary to ensure tree level unitarity at high
energies using model-dependent dipole form factors for all the
couplings, e.g.
$$\Delta\kappa(\hat s) =
\frac{\Delta\kappa}{(1 + \hat s / \Lambda_{\mathrm FF}^2)^2}$$
\noindent
where $\Delta\kappa  =$~value of coupling parameter at $\hat s = 0$,
$\hat s =$~square of the invariant mass of the partonic subprocess,
and $\Lambda_{\mathrm FF} =$~form factor scale, typically taken to be about 2~TeV.
Limits have been obtained by CDF and \d0\ using \wg\ 
production\hspace{4pt}\cite{wg},
and the processes 
$WW \to \ell \nu \ell^\prime \nu^\prime$\hspace{4pt}\cite{wwdilep}, and
$WW/WZ \to \ell \nu jj / \ell^+ \ell^- jj$\hspace{4pt}\cite{wwlvjj}.
A review of these results is given in Ref.~13.
In the following
subsection we report on recent limits derived by \d0\ using a combination of 
these data\hspace{4pt}\cite{d0_simfit_prd}.

\subsection{\d0\ Combined Analysis of \wwg\ and \wwz\ Couplings}

\d0\ has recently performed a simultaneous fit to the photon $p_T$
distribution in the $W\gamma$ data, the lepton $p_T$ distribution in
the $WW \to \ell \nu \ell^\prime \nu^\prime$ data, and the $p_T^{e
\nu}$ distribution in the $WW/WZ \to e \nu jj$ 
data. 
Limits on the \wwg\ and \wwz\ coupling parameters are extracted from
the fit, taking care to account for correlations between the
uncertainties on the integrated luminosity, the selection efficiencies,
and the background estimates.
The results are given in Table~\ref{table:combined}.

The \d0\ fit has also been performed using the alternative parametrization 
of the couplings used by the LEP groups, in terms of the parameters
$\alpha_{B\phi}, \alpha_{W\phi}$, and $\alpha_W$\footnote{
These are related to the previous set by
\begin{eqnarray*}
\Delta g_1^Z = \qwe_{W \phi} / {\mathrm cos}^2\theta_W &\qquad&
\lambda_\gamma = \lambda_Z=\qwe_W \cr
\Delta \kappa_\gamma = \qwe_{W \phi } + \qwe_{ B \phi} &\qquad&
\Delta \kappa_Z = \qwe_{W \phi } - {\mathrm tan}^2\theta_W \qwe_{ B \phi}, 
\end{eqnarray*}
where $\theta_W$ is the weak mixing
angle.
}.
The results are shown in
Table~\ref{table:combined-alpha}. Also, shown are the limits obtained by
combining with the LEP limits reported at this
conference\hspace{4pt}\cite{LEP_wwvlimits}.
Note that the LEP limits should be multiplied by a factor $(1 + s /
\Lambda^2_{\mathrm FF})^2$ to compare directly with the
\d0\ results. At the LEP energy, $\sqrt{s} = 183$~GeV, 
this factor is only 1.017 for $\Lambda_{\mathrm FF} = 2$~TeV. Since this is a
negligible correction, it was not taken into account.

The LEP limits are based on approximately 55~pb$^{-1}$ of data per 
experiment at $\sqrt{s} = 183$~GeV. They are
complimentary to the Tevatron limits because they
are obtained from a different process (i.e. $e^+ e^- \to W^+W^-$)
using angular distributions of the decay products.

\begin{table*}[ht]
{\footnotesize
\begin{center}
\begin{tabular}{ccc}  \hline \hline
Coupling	&$\Lambda_{\mathrm FF}=1.5$~TeV	&$\Lambda_{\mathrm FF}=2.0$~TeV  \\ \hline



$\Delta\kappa_{\gamma}$		&$-$0.63, 0.75   &$-$0.59, 0.72\\

$\lambda_{\gamma}$   		&$-$0.27, 0.25   &$-$0.26, 0.24\\

$\Delta\kappa_Z$		&$-$0.46, 0.64   &$-$0.42, 0.59\\

$\lambda_Z$			&$-$0.33, 0.37   &$-$0.31, 0.34\\

$\Delta g^Z_1$ 		  	&$-$0.56, 0.86   &$-$0.52, 0.78\\

\multicolumn{3}{l} {
Assuming $\kappa_\gamma = \kappa_Z = \kappa$, 
$\lambda_\gamma = \lambda_Z = \lambda$:}\\

$\Delta\kappa$ 			&$-$0.33, 0.46   &$-$0.30, 0.43\\ 

$\lambda$			&$-$0.21, 0.21   &$-$0.20, 0.20\\

\hline 
\end{tabular}
\end{center}
}
\caption{
\d0\ limits on anomalous couplings at the 95\% CL from a 
   simultaneous fit to the $W\gamma$, $WW \to \ell \nu \ell^\prime
   \nu^\prime$, and $WW/WZ \to e \nu jj$ data.}
\label{table:combined}
\end{table*}
\begin{table*}[ht]
{\footnotesize
\begin{center}
\begin{tabular}{cccc}  \hline \hline
Coupling   &\d0\  & LEP combined &\d0\ + LEP combined \\ \hline

$\alpha_{B\phi}$  & $-$0.77, 0.58  & -0.44, 0.95 & $-$0.42, 0.43\\

$\alpha_{W\phi}$  & $-$0.22, 0.44  & -0.12, 0.13 & $-$0.14, 0.10\\

$\alpha_W$	  & $-$0.20, 0.20  & -0.21, 0.27 & $-$0.18, 0.13\\

\hline
\end{tabular}
\end{center}
}
\caption{
\d0\ limits on anomalous couplings $\alpha_{B\phi}$, 
$\alpha_{W\phi}$, $\alpha_W$, and $\Delta g^Z_1$ at the 95\% CL from a
simultaneous fit to the $W\gamma$, $WW \to \ell \nu \ell^\prime
\nu^\prime$, and $WW/WZ \to e \nu jj$ data. Also shown are the LEP 
limits from a combination of ALEPH, DELPH, L3 and OPAL data,
and the LEP + \d0\ combined limits.}
\label{table:combined-alpha}
\end{table*}

\section{Summary}

The $W$ boson mass has been measured  by CDF and \d0\ using Run Ib data.
The \D0\ result is 
$m_W = 80.44 \pm 0.10~{\mathrm (stat)} \pm 0.07~{\mathrm (syst)}$~GeV$/c^2$,
and the preliminary CDF result is
$m_W = 80.43 \pm 0.10~{\mathrm (stat)} \pm 0.12~{\mathrm (syst)}$~GeV$/c^2$.

Measurements of the trilinear gague boson couplings were reported by
\d0\ using a combined fit to $W\gamma$ data, $WW \to \ell \nu
\ell^\prime \nu^\prime$ data, and $WW/WZ \to e \nu jj$ data.  The \d0\
limits are comparable in sensitivity and complimentary in nature to
the combined results from the four LEP experiments, and \d0\ and LEP
have now produced combined limits.


\section*{References}
\begin{thebibliography}{99}
{\footnotesize

\bibitem{pdg}
Particle Data Group, Barnet RM et al. {\it Phys. Rev.} D54:1 (1996)

\bibitem{d0_wmass_1b}
Abbott B et al (\d0\ Collaboration). {\it Phys. Rev. Lett.} 80:3000 (1998);
Abbott B et al (\d0\ Collaboration). FERMILAB PUB-97-422-E,
to be published in {\it Phys. Rev. D.} (1997)

\bibitem{cdf_wmass_1b}
Lancaster M. ``Measurement of the $W$ Mass in the $W \to \mu \nu$
channel with the CDF detector'', seminar presented at Fermilab, April
4 (1997).

\bibitem{mw_previous}
Alitti J et al (UA2 Collaboration). {\it Phys. Lett.} B276:354 (1992);
Abe F et al (CDF Collaboration). {\it Phys. Rev. Lett.} 65:2243 (1990);
Abe F et al (CDF Collaboration). {\it Phys. Rev. Lett.} 75:11 (1995);
Abachi S et al (\d0\ Collaboration). {\it Phys. Rev. Lett.} 77:3309 (1996)

\bibitem{lep_wmass}
Kjaer N. These proceedings.

\bibitem{top_mass}
Heintz U. Presented at the {\it Aspen Winter Conference on Particle
Physics}, Aspen, Colorado (1998)

\bibitem{dr_sm}
Degrassi G et al. {\it Phys. Lett.} B418:209 (1998);
Degrassi G, Gambino P, Sirlin A. {\it Phys. Lett.} B394:188 (1997)

\bibitem{dr_mssm}
Chanowski P et al. {\it Nucl. Phys.} B417:101 (1994);
Garcia D, Sola J. {\it Mod. Phys. Lett.} A9:211 (1994);
Dabelstein A, Hollik W, Mosle W. In {\it ``Perspectives for Electroweak
Interactions in $e^+ e^-$ Collisions''}, ed. Kniehl BA, World Scientific,
Singapore (1995) p345;
Pierce et al. {\it Nucl. Phys.} B491:3 (1997)


\bibitem{wwv_lagrangian}
Hagiwara K, Peccei RD, Zeppenfeld D. {\it Nucl. Phys.} B282:253 (1987)

\bibitem{wg}
Abe~F, et al (CDF Collaboration). {\it Phys. Rev. Lett.} 74:1936 (1995);
Abachi~S, et al (\d0\ Collaboration). {\it Phys. Rev. Lett.} 75:1034 (1995);
Abachi~S, et al (\d0\ Collaboration). {\it Phys. Rev.} D56:6742 (1997);
Abachi~S, et al (\d0\ Collaboration). {\it Phys. Rev. Lett.} 78:3634 (1997);
Benjamin~D. In {\it ``Topical Workshop Proton-Antiproton
Collider Phys., 10th},'' Batavia, Illinois. AIP Conference Proceedings 357,
ed. Raja R, Yoh J, p.370. (1996)

\bibitem{wwdilep}
Abachi~S, et al (\d0\ Collaboration). {\it Phys. Rev. Lett.} 75:1023 (1995);
Abe~F, et al (CDF Collaboration). {\it Phys. Rev. Lett.} 78:4536 (1997);
Abbott~B, et al (\d0\ Collaboration).  FERMILAB PUB-98-076-E,
submitted to {\it Phys. Rev. D.} (1998)

\bibitem{wwlvjj}
Abe~F, et al (CDF Collaboration). {\it Phys. Rev. Lett.} 75:1017 (1995);
Abachi~S et al (\d0\ Collaboration). {\it Phys. Rev. Lett.} 77:3303 (1996);
Abachi~S, et al (\d0\ Collaboration). {\it Phys. Rev. Lett.} 79:1441 (1997);
Nodulman~LJ. In {\it Int. Conf. High Energy Phys., 28th}, Warsaw,
Poland (1996)


\bibitem{arnps_review}
Ellison J, Wudka J. {\it ``Study of Trilinear Gauge Boson
Couplings at the Tevatron Collider''}, to be published in {Ann. Rev. Nucl. 
Part. Sci.}, UCR/{D\O}/98-01, hep-ph/9804322 (1998)

\bibitem{d0_simfit_prd}
Abbott~B, et al (\d0\ Collaboration). FERMILAB PUB-98-094-E,
to be published in {\it Phys. Rev. D.} (1998);

\bibitem{LEP_wwvlimits}
Jousset J-M. These proceedings;\\
Combination of the \d0\ and LEP
results is decribed in {\it ``A Combination of Preliminary Measurements of
Triple Gauge Boson Coupling Parameters Measured by the LEP and \d0\
Experiments''}, \d0\ Note 3437 (1998).

}
\end{thebibliography}
\end{document}
